\documentclass[submission, Phys]{SciPost}

\usepackage{doi}
\usepackage{lineno}
\definecolor{lightblue}{rgb}{0.68, 0.85, 0.9}
\definecolor{airforceblue}{rgb}{0.36, 0.54, 0.66}

\usepackage{comment}

\usepackage{amsmath,amssymb,amsfonts,amsthm}
\usepackage{enumerate}
\usepackage{enumitem}
\usepackage{latexsym}
\usepackage{xcolor}

\newcommand{\bite}{\begin{itemize}}
\newcommand{\eat}{\end{itemize}}
\newcommand{\beq}{\begin{equation}}
\newcommand{\eeq}{\end{equation}}

\newcommand{\beqa}{\begin{align}}
\newcommand{\eeqa}{\end{align}}
\newcommand{\barr}{\begin{array}}
\newcommand{\earr}{\end{array}}

\newcommand{\mc}[1]{\mathcal{#1}}
\newcommand{\mbb}[1]{\mathbb{#1}}
\newcommand{\mf}[1]{\mathfrak{#1}}

\newcommand{\expect}[1]{\langle #1\rangle}

\begin{document}

\begin{center}{\Large \textbf{
Connecting Loop Quantum Gravity and String Theory via Quantum Geometry
}}\end{center}

\begin{center}
D. Vaid\textsuperscript{*}
\end{center}

\begin{center}
Department of Physics, National Institute of Technology Karnataka (NITK), India \\
* dvaid79@gmail.com
\end{center}

\begin{center}
\today
\end{center}


\section*{Abstract}
{\bf
We argue that String Theory and Loop Quantum Gravity can be thought of as describing different regimes of a single unified theory of quantum gravity. LQG can be thought of as providing the pre-geometric exoskeleton out of which macroscopic geometry emerges and String Theory then becomes the \emph{effective} theory which describes the dynamics of that exoskeleton. The core of the argument rests on the claim that the Nambu-Goto action of String Theory can be viewed as the expectation value of the LQG area operator evaluated on the string worldsheet. A concrete result is that the string tension of String Theory and the Barbero-Immirzi parameter of LQG turn out to be proportional to each other.
}

\vspace{10pt}
\noindent\rule{\textwidth}{1pt}
\tableofcontents\thispagestyle{fancy}
\noindent\rule{\textwidth}{1pt}
\vspace{10pt}

\section{Strings vs. Loops}\label{sec:comparison}

There are several competing candidates for a theory of quantum gravity. Two of the strongest contenders are Loop Quantum Gravity (LQG) \cite{Ashtekar1991Lectures, Rovelli2011Zakopane, Ashtekar2004Background} and String Theory \cite{Tong2010Lectures,Zwiebach2009A-First,Polchinski1998aString}. Of these, string theory has been around for much longer, is more mature and has a far greater number of practitioners. LQG is younger, with fewer adherents, but still with the potential to present a serious challenge to the supremacy of String Theory.

There exist critiques of both approaches. In \cite[Sec. 4, pg 6]{t-Hooft2016Natures}, 't Hooft refers to the related approaches of causal sets, causal dynamical triangulations and LQG, as ``\emph{wild concoctions}''. From the LQG side, Lee Smolin has this to say \cite{Smolin2006The-trouble}: \emph{``some string theorists prefer to believe that string theory is too arcane to be understood by human beings, rather than consider the possibility that it might just be wrong''}. Of course, these statements do not necessarily reflect that present views of either of these researchers or of the research community at large, but they do give a general idea of the gulf that divides the two communities.

In this note we would like to suggest that rather than an ``either/or'' situation, one can instead have the best of both worlds. The commonalities between the two approaches are far greater than their apparent differences and that both frameworks provide essential conceptual constructs that will go into any final theory of quantum gravity. The following observations hold true for both fields:

\begin{enumerate}
	\item \textbf{fundamental degrees of freedom are the same} - extended one-dimensional objects referred to as ``strings'' by string theorists and as ``holonomies'' by LQG-ists.
	\item \textbf{identical predictions \cite{Kaul2012Entropy,Sen2012Logarithmic} for the Bekenstein-Hawking entropy of black holes} are obtained\footnote{Though note \cite{Sen2012Logarithmic,Sen2014Microscopic} where disagreements between the two sets of calculations are pointed out. Though certain factors are not the same, the overall form of the entropy area relation including logarithmic corrections is the same. The differences could possibly be traced to the use of Euclidean geometry to determine black hole entropy in string theory. This introduces ingredients which are missing in the LQG calculation.}, albeit both fields follow different routes to get there.
	\item \textbf{geometry becomes discrete, or more appropriately, ``quantized'', as one approaches the Planck scale} and the continuum approximation of a smooth background spacetime breaks down. In string theory this happens because closed strings cannot shrink to zero size due to quantum fluctuations. In LQG there is a natural and explicit construction \cite{Ashtekar1992Weaving,Rovelli1993Area,Rovelli1994Discreteness} of quantum operators for area and volume which have a discrete spectrum and whose minimum eigenvalues are greater than zero.
	\item \textbf{coherent, consistent description of matter degrees of freedom is missing} in both LQG and string theory. Though one can always add matter \emph{by hand} to either theory, it would be much more satisfying if the particles of the Standard Model were to arise naturally as geometrical objects which are part and parcel of either theory. Such proposals do exist \cite{Bilson-Thompson2005A-topological,Bilson-Thompson2006Quantum,Wan2007Braid, Vaid2010Embedding} though they are as yet not successful in explaining the full phenomenology and spectrum of the Standard Model. Remarkably in both LQG and string theory, the basic notions are identical. The end points of spin-network edges in LQG \cite{Morales-Tecotl1994Fermions} or open strings with ends attached to two different D-branes in string theory \cite[Ch. 21]{Zwiebach2009A-First}, play the role of fermions.
	\item \textbf{state space of LQG and string theory can be mapped onto each other}. In LQG the kinematical Hilbert space consists of graphs with edges labeled by representations of $ SU(2) $ and edges labeled by invariant $ SU(2) $ tensors. As it so happens, string theory also contains very similar structures which go by the name of ``string networks''. This similarity was alluded to in an older paper by Ashoke Sen \cite{Sen1997String}:
	\begin{quote}
		\emph{in future a manifestly SL(2,Z) invariant non-perturbative formulation of string theory may be made possible by regarding the string network, instead of string loops, as fundamental objects. This would be similar in spirit to recent developments in canonical quantum gravity, in which loops have been replaced by spin networks.}
	\end{quote}
\end{enumerate}

One could go on to find more areas of agreement between LQG and string theory (and also several areas of disagreement) but, hopefully, these five points should provide sufficient justification to expend some effort in exploring what form a unified framework, which incorporates the essential insights of both LQG and String Theory, might take.

Apart from the five points outlined above, there is a very crucial aspect which connects both string theory and LQG. This has to do with the definition of the action associated with a string in string theory and the quantum operator for area which arises in LQG and will form the core of our argument for the existence of a clear connection between LQG and string theory.

There exist several works in the literature \cite{Thiemann2004The-LQG-String:,Gambini2014Emergence,Bodendorfer2015A-note,Cai2017The-String,Smolin1998Strings,Freidel2017Loop,Zuo2016A-note,Zuo2017Simplicity} which have explored the question of a possible relationship between LQG and String Theory at varying levels of rigor. The recent work \cite{Huggett2017The-Atemporal} in particular explores the question of the emergence of spacetime from both String Theory and LQG.

The plan of this work is as follows. In \autoref{sec:area-operator} we provide a lightning introduction to LQG and introduce the area operator. In \autoref{sec:string-geometry} we discuss how the low-energy effective field theory emerges from the string action and what this entails for the relation between quantum geometry and string theory. In \autoref{sec:pregeom-2} we argue that the Nambu-Goto string action can be see as arising from quantum geometric constructs which are present in LQG and finally in \autoref{sec:discussion} we conclude with some thoughts on the present situation and future developments.

\section{LQG Area Operator}\label{sec:area-operator}

In LQG the basic dynamical variables are \cite{Ashtekar2004Background} a $ \mf{su}(2) $ connection $ A_a^i $ (where $ a,b,c $ three dimensional spatial indices and $ i,j,k $ are Lie algebra indices) and the triad  $ e^a_i $ (which determines the three dimensional metric $ h_{ab} $, of the 3-manifold $ \Sigma $, via the relation: $ h_{ab} = e_a^i e_b^j \delta_{ij} $). These satisfy the Poisson bracket:
\begin{equation}\label{eqn:poisson-bracket}
	\left\{ A_a^i, e^b_j \right\} = \kappa \delta^b_a \delta^i_j
\end{equation}
where $ \kappa = 8 \pi G_N $. Associated with each of these variables one can construct operators which correspond to gauge invariant observables. The connection can be smeared along one-dimensional curves to obtain holonomies, the trace of which is gauge invariant. Holonomies are nothing more than the Wilson loops of field theory. For instance, given the curve $ \gamma $ embedded in the spatial 3-manifold, the holonomy of $ A_a^i $ along $ \gamma $ is given by:
\begin{equation}\label{eqn:holonomy}
	g_\gamma[A] = \mc{P} \exp \left\{ i \alpha \int_{\gamma_0}^{\gamma_1} n^a(x) A_a^i(x) \tau_i dx  \right\}
\end{equation}
where $ x $ is an affine parameter along the curve, $ \mc{P} $ is the path-ordered exponential, $ [\gamma_0, \gamma_1] $ are the start and end points of the curve, $ n^a $ is the unit tangent vector to the curve at $ x $, and $ \tau_i $ are $ 1/2 $ of the Pauli matrices $ \tau_i = \sigma_i/2 $. $ g_\gamma[A] $ is a $ SU(2) $ matrix, whose trace then gives us a gauge-invariant object. $ \alpha $ is the gauge coupling constant.

Similarly the triad fields are smeared over two-dimensional surfaces to obtain the generalized momentum variables:
\begin{equation}\label{eqn:triad}
P(S, f) = \int_S f_i~e^a_j~e^b_k~\epsilon^{ijk}~dx^a dx^b
\end{equation}
where $ S $ is an arbitrary two-dimensional surface embedded in $ \Sigma $ and $ f_i $ is a $ \mf{su}(2) $ valued ``''smearing'' function defined on $ S $. It turns out that using these observables it is possible to construct an operator acting on the kinematical Hilbert space which measures the areas of two-dimensional surfaces. The area of a two-dimensional surface $ S $ with intrinsic metric $ h_{AB} $ is given by:
\begin{equation}\label{eqn:area-2d}
	A_S = \int d^2 x \sqrt{\text{det}(h_{AB})}
\end{equation}
Divide $ S $ into $ N $ cells $ S_I $, with $ I = 1,2,\ldots,N $. Then $ A_S $ can be approximated by:
\begin{equation}\label{eqn:area-2d-approx}
	A_S = \sum_{I=1}^N \sqrt{\text{det}(h^I_{AB})}
\end{equation}
where $ h^I_{AB} $ is the two-dimensional metric in the $ I^\text{th} $ cell. Now as $ N $ increases in \eqref{eqn:area-2d-approx}, $ A_S $ will become a better and better approximation to the actual area of $ S $. Of course, it should be kept in mind that when working in the regime where a classical geometry is yet to emerge, the ``actual'' or ``exact'' area of a surface is not a well-defined quantity.

To evaluate the determinant we use the following expression for the determinant of a rank $ n $ matrix:
\begin{equation}\label{eqn:determinant}
	\text{det} A = \sum_{i_1, i_2, \ldots,i_n} \epsilon^{i_1 i_2 \ldots i_n} A_{1 i_1} A_{2 i_2}\ldots A_{n i_n}
\end{equation}
The determinant of $ h_{AB} $ is thus:
\begin{equation}\label{eqn:determinant-2d}
	\text{det} (h) = \sum_{i_1,i_2} \epsilon^{i_1 i_2} h_{1 i_1} h_{2 i_2}
\end{equation}
Now let us setup a local flat co-ordinate system with unti orthogonal basis vectors $ \{\vec{e}_x, \vec{e}_y, \vec{e}_z\} $, such that $ \{ \vec{e}_x, \vec{e}_y \} $ span the tangent space of $ S_I $, while $ \vec{e}_z $ is normal to $ S_I $. Then evaluating the r.h.s of \eqref{eqn:determinant-2d} explicitly and using the fact that $ \epsilon_{ijk} e_x^j e_y^k = e_z^i $, we obtain:
\begin{equation}\label{area-2d-approx-b}
	A_S = \sum_{I=1}^N \sqrt{\vec{e}^I_z \cdot \vec{e}^I_z}
\end{equation}
where $ \vec{e}^I_z $ is the unit normal basis vector to the surface elements $ S_I $. Now, in order to construct the \emph{quantum} operator corresponding to the classical variable $ A_S $, we promote the Poisson bracket \eqref{eqn:poisson-bracket} to a commutator:
\begin{equation}\label{eqn:lqg-commutator}
[ \hat A_a^i, \hat e^b_j ] = i \hbar \kappa \delta^b_a \delta^i_j
\end{equation}
which implies that the triad operator can be written in terms of the connection variable as:
\begin{equation}\label{eqn:triad-operator}
	\hat e_a^j = -i\hbar \kappa \frac{\delta}{\delta A^a_j}
\end{equation}
Substituting this expression for the classical variable in \eqref{area-2d-approx-b} we finally obtain the expression for the quantum area operator:
\begin{equation}\label{eqn:area-quantum}
	\hat A_S = \hbar \kappa \sum_{I=1}^N \sqrt{ - \delta_{jk} \frac{\delta}{\delta A^z_j} \frac{\delta}{\delta A^z_k}} 
\end{equation}

\begin{figure}[tbph]
\centering
\includegraphics[width=0.4\linewidth]{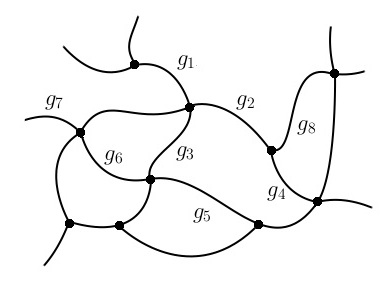}
\caption{A graph state $ \Psi_\Gamma $ consists of an arbitrary graph $ \Gamma $, along with an assigment of an element of $ SU(2) $ (a holonomy) to each edge.}
\label{fig:graph-state}
\end{figure}

\begin{figure}[tbph]
\centering
\includegraphics[width=0.4\linewidth]{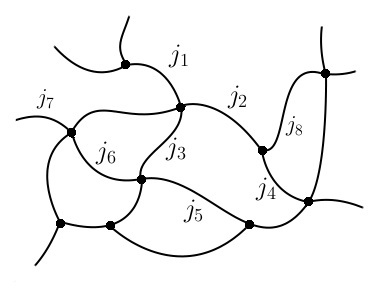}
\caption{The action of the area operator on the graph state $ \Psi_\Gamma $, leads to a labeling of each edge by an eigenvalue of the $ \mf{su}(2) $ Casimir, which are nothing but spin labels labeled by half-integers $ j_i \in \mbb{Z}^+/2 $. In this representation, the graph state is referred to as a ``spin-network''}
\label{fig:spin-state}
\end{figure}

Now in order to find the eigenvalues of this operator we have to act with it upon a state of quantum geometry. Such states are functionals of the form:
\begin{equation}\label{eqn:graph-state}
	\Psi_\Gamma = \psi(g_1,g_2,\ldots,g_n)
\end{equation}
where $ \Gamma $ is an arbitrary graph embedded in $ \Sigma $, $ g_1,\ldots,g_n $ are the holonomies \eqref{eqn:holonomy} of the connection evaluated along each edge of $ \Gamma $ (c.f. \autoref{fig:graph-state}). The action of the momentum operator on the holonomy $ g_\gamma[A] $ is given by:
\begin{equation}\label{eqn:edge-momentum}
	\frac{\delta}{\delta A^a_i} g_\gamma[A] = n_a (x) \tau^i g_\gamma[A]
\end{equation}
and therefore:
\begin{equation}\label{eqn:graph-momentum}
	\frac{\delta}{\delta A^a_i(x)} \Psi(g_1,\ldots,g_k,\ldots,g_n) = n_a^k(x) \tau^i \Psi
\end{equation}
where $ n_a^k(x) $ is the tangent vector to the $ k^\text{th} $ edge at the point $ x $. Consequently the action of the area operator \eqref{eqn:area-quantum} on the graph state $ \Psi_\Gamma $ can be easily seen to be:
\begin{align}\label{eqn:area-quantum-2}
	\hat A_S \Psi_\Gamma & = \hbar \kappa \sum_{I=1}^N \sqrt{-\delta_{jk} \alpha^2 n_a n_a \tau^j \tau^k } \Psi_\Gamma \nonumber \\
	& = \hbar \kappa \sum_{I=1}^N \sqrt{- \tau^2} \Psi_\Gamma
\end{align}
where in the first line it is understood that the sum on the r.h.s is only over those cells of $ S $ which are pierced by an edge of the graph $ \Gamma $. In the second line we have used the fact that tangent vector is normalized: $ n^a n_a = 1 $. Recalling that $ \alpha = 8 \pi G_N $, we can write \eqref{eqn:area-quantum-2} as:
\begin{equation}\label{eqn:area-quantum-3}
	\hat A_S \Psi_\Gamma = 8 \pi l_{PL}^2 \sum_k \sqrt{ j_k (j_k + 1)} \Psi_\Gamma
\end{equation}
where the Planck length $ l_{PL} = \sqrt{\hbar G_N} $ (if $ c = 1 $), the sum is over all the points where edges of $ \Gamma $ intersect $ S $ and $ j_k $ is the eigenvalue of the Casimir operator $ \tau^2 $ along that edge (c.f. \autoref{fig:spin-state})

The same operator can also be realized by starting from the expression for the triad observable $ P(S,f) $ in \eqref{eqn:triad} as follows \cite[Sec 5.1.1]{Ashtekar2004Background}:
\begin{equation}\label{eqn:area-observable}
	A_S = \sum_{I=1}^{N}\sqrt{P^i(S_I) P^j(S_I) \eta_{ij}}
\end{equation}
where $ P^i(S_I) $ is the quantity in \eqref{eqn:triad} with the integral evaluated on the $ I^\text{th} $ cell, with $ f_i = \tau_i $ and constant triad $ e^a_i $. Evaluating \eqref{eqn:triad} on $ S_I $ we find:
\begin{equation}\label{eqn:triad-2}
	P^i = \epsilon_{ijk} e_x^j e_y^k = e_z^i
\end{equation}
where we have used that fact that the cross product of any two elements of a unit orthogonal triad yields the third element. Then $ A_S $, restricted to the $ I^\text{th} $ cell becomes:
\begin{equation}\label{eqn:area-2}
	A_{S_I} = \sqrt{ \vec{e}_z \cdot \vec{e}_z }
\end{equation}
which is just the result \eqref{area-2d-approx-b} obtained previously. From this point the manipulations are as outlined after \eqref{area-2d-approx-b}.

\subsection{Minimum Area and Conformal Symmetry}\label{sec:conformal}

\begin{figure}[tbph]
\centering
\includegraphics[width=0.4\linewidth]{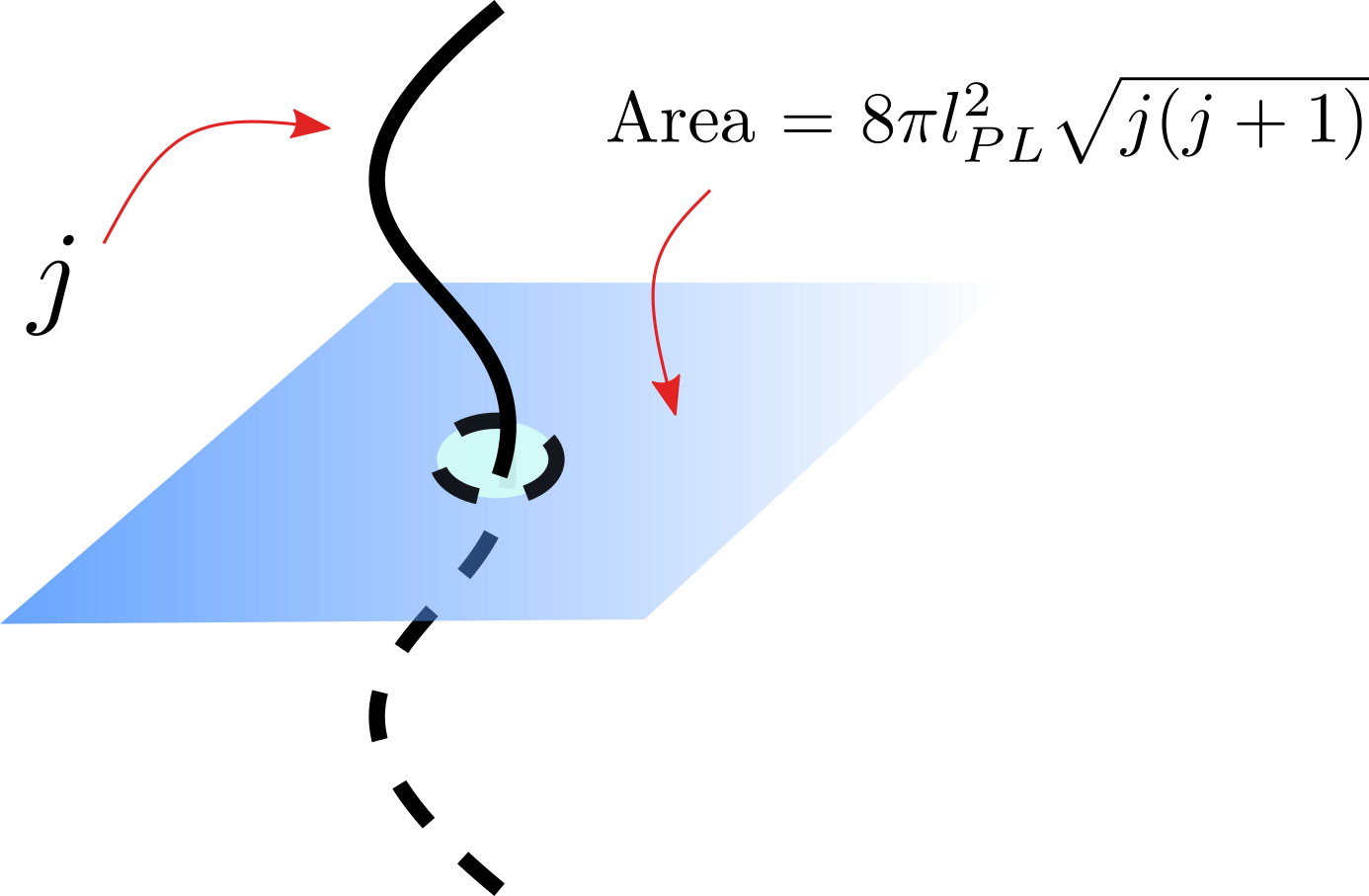}
\caption{A single edge of a graph state punctures a surface and endows it with an area $ 8\pi l^2_{PL} \sqrt{j (j + 1)} $, where $ j $ is the $ SU(2) $ spin representation carried by the given edge.}
\label{fig:area-puncture}
\end{figure}

The quantum operator corresponding to $ A_S $ turns out to possess a minimum eigenvalue:
\begin{equation}\label{eqn:area-eigenvalue}
	A_{S_I} = 8\pi l^2_{PL} \sqrt{j_I (j_I + 1)}
\end{equation}
where $ j_I $ denotes the representation of $ \mf{su}(2) $ assigned to the $ I^{\text{th}} $ cell of the surface $ S $ (c.f. \autoref{fig:area-puncture}). Since the smallest eigenvalue of the angular momentum operator is $ j = 1/2 $, the minimum quantum of area permitted by LQG is:
\begin{equation}\label{eqn:area-minimum}
	A_{min} = 2 \sqrt{3} \pi l^2_{PL} 
\end{equation}

This has an immediate implication for any field theories living on two-dimensional surfaces - the presence of a length scale implies that in such theories conformal symmetry will always be broken. We will see a bit later what this might imply when viewed from the perspective of string theory.

\section{String Theory and Quantum Geometry}\label{sec:string-geometry}

Now, recall that the Nambu-Goto action for a $ 1+1 $ dimensional string worldsheet $ \Sigma $ embedded in some $ D+1 $ dimensional \emph{worldvolume} $ \mc{M} $ is given by \cite{t-Hooft2004Introduction,Tong2010Lectures,Zwiebach2009A-First}:
\begin{equation}\label{eqn:ng-action}
	S_{NG} = -T \int d\tau d\sigma \sqrt{-\text{det} (h_{AB})}
\end{equation}
where $ T $ is the string tension, $ \tau, \sigma $ are the timelike and spacelike co-ordinates respective on the string worldsheet and $ h_{AB} $ is the two-dimensional metric induced on $ \Sigma $ due to its embedding in $ \mc{M} $. There is only term in this action $ \sqrt{-h_{AB}} $ and that is precisely the area of the string worldsheet (the $ 1+1 $ dimensional surface the string sweeps out as it evolves in spacetime). The system that \eqref{eqn:ng-action} represents is a single string evolving in a flat background spacetime.

\subsection{Gravity From String Theory}

Of course, just as the action for a single free particle cannot capture the complexity of a manybody system, the action for a single free string is unlikely to capture the complexity of gravitational and particle physics. Now, we know that the spectrum of a free string embedded in $ D $ spacetime dimensions (the ``worldvolume'') contains, in addition to an infinite tower of left and right moving massive modes, three massless fields described respectively by a traceless symmetric tensor $ h_{\mu\nu}(X) $, an anti-symmetric tensor $ B_{\mu\nu}(X) $ and a scalar $ \Phi(X) $. These objects are identified with, respectively, a graviton, a Kalb-Ramond field and a dilaton \cite[Sec 2.3.2]{Tong2010Lectures}. Here $ \mu, \nu $ are worldvolume indices and $ X^{\mu}(\tau,\sigma) $ are worldvolume co-ordinates which describe the embedding of the string in the worldvolume. The $ \{X^\mu\} $ can also be thought of as scalar fields living on the string worldsheet. This can be seen by writing the string action in the form associated with Polyakov's name:
\begin{equation}\label{eqn:polyakov-flat}
	S_{Polya} = -\frac{T}{2} \int d\tau d\sigma \sqrt{-g} g^{ab} \partial_a X^\mu \partial_b X^\nu \eta_{\mu\nu}
\end{equation}
where $ a,b \in (1,2) $ are worldsheet indices, $ g_{ab} $ is the intrinsic metric of the worldsheet and $ \eta_{\mu\nu} $ is the flat worldvolume metric. In this form it is clear that the string action can be thought of describing $ D+1 $ massless, non-interacting scalar fields living on a two-dimensional manifold.

The traceless symmetric tensor $ h_{\mu\nu}(X) $ has spin 2 and is therefore the source of the claim that gravity is already present in string theory. The argument \cite{Boulware1975Classical} is that the only consistent interacting theory that can be constructed from a spin-2 field has to be General Relativity\footnote{This does not, however, appear to be an entirely settled point \cite{Padmanabhan2004From,Deser2009Gravity}.}. An alternate route to obtain the low-energy effective action is to replace the flat metric $ \eta_{\mu\nu} $ with a general curved metric $ G_{\mu\nu} $ in \eqref{eqn:polyakov-flat}:
\begin{equation}\label{eqn:polyakov-gravity}
		S_{Polya} = -\frac{T}{2} \int d\tau d\sigma \sqrt{-g} g^{ab} \partial_a X^\mu \partial_b X^\nu G_{\mu\nu}(X)
\end{equation}
The requirement that \eqref{eqn:polyakov-gravity} satisfy Weyl invariance implies \cite[Sec 3.7]{Polchinski1998aString}, \cite[Sec 7.2]{Tong2010Lectures} that the beta functions for the graviton, the antisymmetric tensor and the dilaton, must be zero. This in turn implies that the background in which the string is propagating must satisfy Einstein's field equations with source terms coming from the antisymmetric tensor and the dilaton. For example, the low energy effective action for the bosonic string (in $ D=26 $ spacetime dimensions) takes the form \cite{Tong2010Lectures}:
\begin{equation}\label{eqn:effective-action}
	S_{eff} = \frac{1}{2 \kappa^2_0} \int d^{26} X \sqrt{-G} e^{-2\Phi} \left( R - \frac{1}{12} H_{\mu\nu\lambda}H^{\mu\nu\lambda} + 4 \partial_\mu \Phi \partial^\mu \Phi \right)
\end{equation}
where $ H_{\mu\nu\lambda} $ is the curvature of the antisymmetric tensor field $ B_{\mu\nu} $.

Let us pause for a moment to appreciate what a remarkable result this is. Starting from nothing but a free string propagating in a flat spacetime, the requirement of worldsheet Weyl invariance implies that the background geometry must necessarily satisfy Einstein's equations. As David Tong puts it so eloquently \cite[pg. 175]{Tong2010Lectures}:
\begin{quote}
	\emph{That tiny string is seriously high-maintenance: its requirements are so stringent that they govern the way the whole universe moves.}
\end{quote}
We do not wish to quibble with the validity of the arguments which lead from \eqref{eqn:polyakov-flat} to \eqref{eqn:polyakov-gravity} to \eqref{eqn:effective-action}. There is however a very important point which we have overlooked in this discussion and which undermines the claim that ``string theory contains gravity''. To understand this point we must think like a LQG-ist.

\subsection{Geometry vs. Pre-Geometry}\label{sec:pregeom-1}

While the elegance and power of the string theoretical arguments cannot be disputed, seen from the perspective of LQG, string theory has a fundamental flaw. LQG leads us to a concrete notion of \emph{quantum geometry}. It allows us to construct a framework in which we can do away with the notion of classical geometries - whether flat or curved - entirely, and instead starting with the \emph{atoms} of spacetime \cite{Krasnov2009Black,Rovelli2006A-semiclassical} using which we can build up almost any kind of geometry we can think of\footnote{Though not all such geometries will be stable against perturbations. The formalism of Causal Dynamical Triangulations (CDT), closely related in spirit to LQG allows one to study this question in detail \cite{Ambjorn2013Quantum}.}. Using the terminology of John Wheeler, starting with \emph{pre-geometry}, we can construct geometry.

In much the same way that the Pauli principle and the theory of linear combination of atomic orbitals (LCAO) shows us that atoms can be brought together to form molecules only in certain combinations, the kinematical constraints of quantum geometry determine how the atoms of quantum geometry can be ``glued'' together to form more complicated structures. Thus, in LQG, we make no presumption of a classical background spacetime. Not only is LQG a theory of quantum gravity which does not depend on the background geometry, it is a theory \emph{in which there is no background} to begin with! This is in sharp contrast to string theory, where \emph{in the very first step itself} \eqref{eqn:polyakov-flat} we assume that there exists a smooth, flat background spacetime on which the string can propagate. While it \emph{is} true, that starting from a string in a flat background, theoretical consistency ultimately requires the background to ultimately satisfy Einstein's equations, this does not obviate the fact that the existence of smooth, continuum background spacetime is taken for granted in conventional formulations of string theory.

It seems clear that any theory which claims to be a theory of ``quantum gravity'' must explain how spacetime arises in the first place rather than putting it in by hand at the very beginning. String theory fails this test. However, and this is crucial, this fact does not invalidate the results obtained in string theory. It only motivates us to try and understand whether the tools of LQG can be harnessed to provide the missing link between pre-geometry and string theory.

\section{Geometry from Pre-Geometry}\label{sec:pregeom-2}

Let us make the assumption that, \emph{a priori}, the spacetime manifold has no structure, no metric, no way to measure distances and areas. In that case the question arises as to how are we to define the integrand of the Nambu-Goto action in \eqref{eqn:ng-action}. The clue lies in LQG. Recall that in \autoref{sec:area-operator}, we explained how the edges of a graph state in LQG carry angular momenta and how these angular momenta endow surfaces pierced by those edges with quanta of area. This leads to the expression \eqref{eqn:area-eigenvalue} for the area of a surface in terms of the angular momenta $ j_i $ carried by each edge which pierces that surface:
\begin{equation}\label{eqn:surface-area}
	A_{S_I} = 8\pi l^2_{PL} \sqrt{j_I (j_I + 1)} \tag{\ref{eqn:area-eigenvalue}}
\end{equation}
Now, imagine that there are many strings moving around in this structureless manifold, with each string carrying some angular momentum in some representation of $ \mf{su}(2) $. Each string traces out a two-dimensional surface - or ``worldsheet'', but as yet without any notion of area - as it moves through the manifold. We would expect that the worldsheets generated by the different strings would inevitably intersect. If seen from the perspective of just a single string, its worldsheet will be pierced by the other strings at various points. According to the LQG prescription \eqref{eqn:area-eigenvalue}, at each such point the worldsheet of the original string will be endowed with one quantum of area determined by the angular momentum carried by the corresponding string. The total area of the worldsheet will then be given by:
\begin{equation}\label{eqn:surface-area-2}
		\expect{\hat A_S} = 8 \pi l_{PL}^2 \sum_k \sqrt{ j_k (j_k + 1)}
\end{equation}
How can we understand string evolution from this perspective? One would expect that strings would tend to avoid running into each other. This expectation can be converted into a mathematical statement by requiring that \eqref{eqn:surface-area-2} be minimized. But this is nothing more than the statement of the extermization of the Nambu-Goto action \eqref{eqn:ng-action} of the string worldsheet! Finally we are left with the following conjecture. The Nambu-Goto action for the bosonic string can be expressed in terms of the expectation value of the LQG area operator acting on a pre-geometric graph state:
\begin{equation}\label{eqn:emergent-action}
	S_{NG} \propto \expect{\Psi\vert \hat A \vert \Psi}
\end{equation}
Notice that we have used proportionality instead of equality in the above expression. This is because, a priori, the two quantities on either side have different units. $ S_{NG} $ has units of action or energy per unit time and the area has units of $ L^2 $ or $ E^{-2} $. In order to equate both sides suitable proportionality constants must be introduced. This will lead us to the relationship between the Barbero-Immirzi parameter $ \beta $ of LQG \cite{Barbero1996From,Immirzi1996Real} and the string tension $ T $ (or alternatively the Regge slope $ \alpha = 1/4\pi T$). This is the topic of the next section.

In passing let us note that one consequence of the quanitzed nature of the string worldsheet is that, as mentioned in \autoref{sec:conformal}, the conformal invariance of the worldsheet would no longer be an exact symmetry. This would have immediate implications for the number of spacetime dimensions in which one could define a consistent string theory.

\subsection{Barbero-Immirzi parameter as String Tension}

In the definition of the classical phase space and therefore of the kinematical Hilbert space of LQG, there is a one-parameter ambiguity \cite{Barbero1996From,Immirzi1996Real} known as the Barbero-Immirzi parameter. Let us first briefly recall its origin.

In the original formulation of Ashtekar \cite{Ashtekar1986New,Ashtekar1987New}, the $ \mf{su}(2) $ connection $ A_a^i $ was obtained by performing a canonical transformation on the Palatini phase space consisting of the extrinsic curvature one-form $ K^i_a = K_{ab} e^{bi} $ and the triad $ e^a_i $:
\begin{equation}\label{eqn:phase-space}
	K^i_a, e^j_b \Rightarrow A^i_a, e^j_b
\end{equation}
where:
\begin{equation}\label{eqn:ashtekar-transform}
	A^i_a = \Gamma^i_a + i K^i_a
\end{equation}
here $ \Gamma^i_a $ is the Levi-Civita connection one-form, which annihilates the triad:
$$ D_a e^b_i = \partial_a e^b_i + \epsilon_{ij}{}^k \Gamma^j_a e^b_k $$.

It was later realized in \cite{Barbero1996From,Immirzi1996Real}, that instead of the unit imaginary $ i $, one could use \emph{any} complex number $ \beta \in \mbb{C}$ in \eqref{eqn:ashtekar-transform}. While all the phases spaces with different values of $ \beta $ are equivalent at the \emph{classical} level, in the \emph{quantum} theory different values of $ \beta $ correspond to physically distinct Hilbert spaces with inequivalent spectra of the fundamental geometric operators. In particular the spectrum of the area operator \eqref{eqn:surface-area} is modified to become:
\begin{equation}\label{eqn:surface-area-immirizi}
	A_{S_I} = 8\pi \beta ~ l^2_{PL} \sqrt{j_I (j_I + 1)} \tag{\ref{eqn:area-eigenvalue}}
\end{equation}
In light of this modification, let us reconsider the two sides of \eqref{eqn:emergent-action}. The left side has the form:
\begin{equation*}
	S_{NG} \simeq tension \times area
\end{equation*}
and the right side:
\begin{equation*}
	\expect{\Psi\vert \hat A \vert \Psi} \simeq \beta \times area
\end{equation*}
Demanding equality of the two sides would therefore imply that the string tension $ T_{string} $ and the Barbero-Immirzi parameter $ \beta $ are related to each other by some, as yet undetermined, constant $ T_{loop} $:
\begin{equation}\label{key}
	T_{string} = \beta ~ T_{loop}
\end{equation}
Since $ \beta $ is dimensionless, the proportionality constant will also have units of tension. Hence the nomenclature $ T_{loop} $ is used to indicate that this quantity is a tension associated with the 1D graph edges of LQG states.

\section{Discussion and Future Work}\label{sec:discussion}

We have presented a rough outline of a method in which one can resolve the central challenges of both LQG and string theory. The problem with LQG is the lack of a clear method to obtain a semiclassical spacetime geometry in the limit of large graph states. The flaw with string theory, in the author's humble opinion, is the presumption of the existence of a background geometry on which the string can propagate. Both of these problems can be cured by viewing the Nambu-Goto action, and the geometry of the string worldsheet, as arising from a pre-geometry which can be described in the language of LQG. From this perspective, String Theory is the glue which connects a quantum theory of geometry (LQG) to a classical theory of gravity with matter such the one given in \eqref{eqn:effective-action}.

One might then be tempted to suggest that, if the arguments presented in this paper hold up to greater scrutiny, it is LQG which is the ``\emph{true}'' theory of quantum gravity, whereas String Theory is merely an ``\emph{effective}'' theory. However, such a claim would again be based on the same sort of hubris which has affected the high energy community in the past. Though LQG might provide the microscopic description of the atoms of quantum geometry, without String Theory one cannot connect this microscopic description with the low energy physics of the world around us. Both are ultimately descriptions of Nature at different scales and without either theory we will not have a proper understanding of quantum gravity.


It is my belief that Nature could not have been so malicious as to show us two avenues of investigation, with both being fully grounded in physical theory and both leading to concrete predictions about the nature of the physical universe, only for us to eventually discover that one or the other of those avenues belonged to the realm of fantasy all along.

Much more work remains to be done to put the ideas of this work on a firm mathematical footing. Several questions present themselves in search of a resolution. We have tried to list some of these questions and  have suggested possible solutions, in the following\footnote{the author is grateful to the referees of the ``Beyond Spacetime'' 2017 essay contest for making these valuable observations}:
\begin{enumerate}
	\item \textbf{LQG areas correspond to spacelike surfaces, whereas the string worldsheet is not spacelike}
	
	while this observation is correct, it is important to note that the string worldsheet coordinates can always be Wick rotated so that instead of a field theory in one space and one time dimension, we can work with a theory in two spatial dimensions. One can thus think of left hand side of \eqref{eqn:emergent-action} as the 2D spatial string action. That being said, one would have to transform back to $ 1+1 $ dimensions to make contact with the usual string theory. It is not clear to me at the moment what the corresponding Wick rotation for the LQG area operators would be.
	
	\item \textbf{What of the string spectrum?}
	
	since the string worldsheet is no longer a continuous surface, we cannot expect the strings themselves to be continuous objects. In this picture a string corresponds to any path connecting two vertices of a LQG graph state. Each such path is the union of edges connecting all the vertices which lie of that path. In the quantum theory each edge is labeled by a representation of $ SU(2) $. Thus, the LQG string (so to speak) $ l $, would be a piecewise continuous object, consisting of segments $ l_i$, labeled by $ \mf{su}(2) $ spins ($ S_i $), joined together by operators $ \mc{O}_{ij} : \mc{H}_i \otimes \mc{H}_j \Rightarrow \mbb{C} $ which map the Hilbert space of neighboring segments into each other: $ \mc{O}_{ij}(S_i, S_j) \in \mbb{C} $. In effect, the LQG string would be a one-dimensional spin system.
	
	\item \textbf{How is a path integral to be computed?}
	
	the path integral represents all possible paths the string can take in going from an initial state to a final state. The resulting integral has a sum over all possible surfaces (worldsheets) which have the initial and final states as their boundaries. In the present case, such an integral would be replaced by a sum over all possible combinations of spins which puncture the worldsheet. Rather than varying the worldsheet we would need to vary the values of the spin-assignments to the graph edges which puncture the worldsheet. The end result, of course, will be the same. Only those configurations which extermize the total area would make the greatest contribution to the resulting sum (or integral).
	
\end{enumerate}

This is by no means a complete list of objections to or criticisms of this proposal, but it is my hope that string theorists, LQG-ists, CDT-ists and all other camps can come together to help construct a complete, consistent theory of quantum gravity so that we might finally be able to understand what lies ``beyond spacetime''.

\textbf{Note:} An earlier version of this paper was submitted as an entry to the ``Beyond Spacetime, 2017'' essay contest on Sep 25, 2017.

\section*{Acknowledgments}

The author would like to dedicate this essay to his loving wife on the occassion of her birthday.

\paragraph{Funding information}

DV wishes to acknowledge the support of a visiting associate fellowship from the Inter-University Centre For Astronomy And Astrophysics (IUCAA), Pune, India, where a portion of this work was completed.

\bibliography{unifying-scipost.bib}

\nolinenumbers

\end{document}